\documentclass[10pt, conference, compsocconf]{IEEEtran}

\usepackage{mathrsfs}
\usepackage{graphicx}
\usepackage{amsfonts}
\usepackage{dsfont}
\usepackage{bbm}
\usepackage{amsmath}
\usepackage{bm} 
\usepackage{url}
\usepackage{url}
\usepackage{amsmath}
\usepackage{perpage} \MakePerPage{footnote}
\ifCLASSINFOpdf

\else
  
\fi

\begin{document}

\title{MatLM: a Matrix Formulation for Probabilistic Language Models}
%

\author{\IEEEauthorblockN{Yanshan Wang}
\IEEEauthorblockA{Dept. of Health Sciences Research\\
Mayo Clinic\\
Rochester, USA\\
Wang.Yanshan@mayo.edu}
\and
\IEEEauthorblockN{Hongfang Liu}
\IEEEauthorblockA{Dept. of Health Sciences Research\\
Mayo Clinic\\
Rochester, USA\\
Liu.Hongfang@mayo.edu}
}

\maketitle

\begin{abstract}
Probabilistic language models are widely used in Information Retrieval (IR) to rank documents by the probability that they generate the query. However, the implementation of the probabilistic representations with programming languages that favor matrix calculations is challenging. In this paper, we utilize matrix representations to reformulate the probabilistic language models. The matrix representation is a superstructure for the probabilistic language models to organize the calculated probabilities and a potential formalism for standardization of language models and for further mathematical analysis. It facilitates implementations by matrix friendly programming languages. In this paper, we consider the matrix formulation of conventional language model with Dirichlet smoothing, and two language models based on Latent Dirichlet Allocation (LDA), i.e., LBDM and LDI. We release a Java software package--MatLM--implementing the proposed models. Code is available at: \url{https://github.com/yanshanwang/JGibbLDA-v.1.0-MatLM}.
\end{abstract}

\begin{IEEEkeywords}
information retrieval; language model; software package; matrix formulation
\end{IEEEkeywords}

\section{Introduction}

Information Retrieval (IR) has become the a crucial part of modern information systems. It is the science of analysis, organization, storage, searching and retrieval of information \cite{salton1986introduction, croft2010search}. IR provides users with desired information from a large collection given a query. The retrieved documents are ranked in descending order according to their relevance to the query. The ranking model is one of the fundamental parts of IR pipeline since it is directly related to the final document list shown to the users. 

A lot of IR models have been proposed amongst which the strength of probabilistic language models have been evaluated by many studies \cite{croft2010search, zhai2001study, blei2003latent, wei2006lda, wang2015indexing}. Users normally need to select a set of query terms that might appear in the relevant documents in order to retrieve the most relevant documents. The language modeling in IR takes advantage of this idea by assuming that a document is an accurate match to a query if the document is likely to generate the query. Query likelihood language model (QLM) is the simplest probabilistic language model that ranks documents by the probability that the query could be generated by the document\cite{croft2010search}. Yet, QLM may encounter data sparsity problem when it is applied to a limited amount of text. In order to overcome the data sparsity problem, smoothing techniques are leveraged by the probabilistic language models. Dirichlet smoothing language model (LMD) is generally considered to be more effective than other smoothing based language models, especially for short queries \cite{zhai2001study}. More sophisticated methods are reported in the literature by addition of latent information to the smoothing based probabilistic language models. Latent Dirichlet Allocation (LDA)-based document model (LBDM) \cite{wei2006lda} and Indexing by LDA (LDI) \cite{wang2015indexing} are two models taking advantage of the latent topical information generated by LDA \cite{blei2003latent}. 


Although the above mentioned methods have been studied in the literature, there are few publicly available packages implementing these methods. Lucene\footnote{Available at: https://lucene.apache.org/} and Lemur\footnote{Available at: http://www.lemurproject.org/} are the only public packages that implement LMD. However, it is intractable to adjust the codes to a new language model due to its implementation within a specific framework. In addition, no packages could be found for LBDM and LDI, to the best of our knowledge. 

The motivation of our study is to reformulate the probabilistic language models in terms of matrix calculation. This will help to speed up the computation by utilizing Graphics Processing Units (GPUs) empowered matrix calculation or simplify the implementation using programming language that favor high-performance matrix algebra such as Matlab, and Python. Moreover, we can directly link the input data stored in vectors, and output data, which are query-document similarity scores, through matrix transformations. This matrix methodology is a potential formalism for standardization of language models and optimization of mathematical parameters \cite{wang2012recovery}. In this study, we report a Java package-MatLM--implementing the matrix formulations of LMD, LBDM and LDI. 

The reminder of the paper is organized as follows. In Section \ref{sec.pr}, we introduce the probabilistic representations of LMD, LBDM and LDI. Section \ref{sec.mr} demonstrates notations and the proposed matrix reformulations. Section \ref{sec.ie} presents the implementation of the proposed method and its performance. Finally we conclude the paper in Section \ref{sec.cd}.

\section{Probabilistic Representations}
\label{sec.pr}
Query Likelihood Language Model (QLM) is the basic approach for ranking documents using probabilistic language models
\begin{equation}
  p(Q|D)=\prod_{q\in Q} p(q|D)
\end{equation}
where $q$ is a query term in query $Q$, and $D$ is a document. Given that query terms are independent, the probability of query given document is calculated by the product of query terms given document. Due to loss of accuracy in binary storage of multiplication of very small numbers, logarithmic scale is used to represent the equation above, i.e.,
\begin{equation}
  \log p(Q|D)=\sum_{q\in Q} \log p(q|D).
\end{equation}
Since the score $p(q|D)$ will be zero if a query word is missing in the document, the Dirichlet smoothing language model(LMD) is utilized to estimate $p(q|D)$. The smoothed probability estimates can be written as \cite{zhai2001study}
\begin{equation}
\label{equ.qlm}
  p(w|D)=\frac{|D|}{|D|+\mu} p'(w|D)+(1-\frac{|D|}{|D|+\mu})p'(w|C), 
\end{equation}
where $w$ denotes the term, $|D|$ the number of terms in the document $D$, and $\mu$ the Dirichlet prior. $p'(w|D)=\frac{n_w}{N_w}$ is the maximum likelihood estimate (MLE) of the term $w$ in the document $D$ where $n_w$ denotes the frequency of word $w$ in document $D$ and $N_w$ denotes the total number of words in document $D$. $p'(w|C)=\frac{n_w^C}{N_w^C}$ is the MLE of the term $w$ in the collection $C$ where $n_w^C$ and $N_w^C$ denote the frequency of word $w$ and the total number of words in the corpus $C$. To calculate the probability of query term given document, we can estimate $p(q|D)$ by finding the query term that occurs in the collection, i.e., $p(q|D)=p(w_q|D)$, for $w_q\in Q$.

By linearly combining the original document model in Equ. \ref{equ.qlm} and latent information given by LDA, the probability function of LBDM is \cite{wei2006lda} 
\begin{equation}
\label{equ.lbdm}
\begin{aligned}
  p(w|D)=&\lambda(\frac{|D|}{|D|+\mu} p'(w|D)+(1-\frac{|D|}{|D|+\mu})p'(w|C)) \\
  & +(1-\lambda)\sum_{z=1}^{K}p(w|z)p(z|D), 
\end{aligned}
\end{equation}
where $\lambda$ is a parameter, $z$ is the latent topic and $K$ the number of topics.

In the LDI model, the term and document probabilities are represented in a topic space \cite{wang2015indexing}
\begin{equation}
 p(z|w)=\frac{p(w|z)p(z)}{\sum_{z}p(w|z)p(z)}, 
\end{equation}
and
\begin{equation}
 p(z|D)=\sum_{w\in D}p(z|w)p(w|D), 
\end{equation}
where $p(w|D)=\frac{n_w}{N_w}$. The ranking scores are the cosine similarities between the query and document probabilities, i.e.,
\begin{equation}
\label{equ.ldi.s}
similarity(Q,D)=\left< p(z|Q), p(z|D) \right>,
\end{equation}
where $p(z|Q)=\sum_{w\in Q}p(z|w)p(w|Q)$ is the query probability in the topic space.

\section{Matrix Reformulations}
\label{sec.mr}

\begin{figure}[b]
  \centering
  \includegraphics[width=0.4\textwidth]{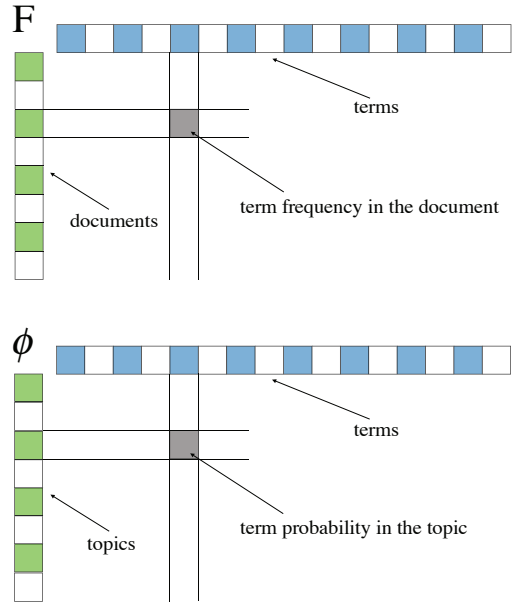}
  \caption{Matrix representations of document-term and topic-term matrices.}
  \label{fig.matrix}
\end{figure}

From the probabilistic language models described in Section \ref{sec.pr}, we observe that the probability calculations are elementwise operations. Since the input data such as term frequencies or topical probabilities can be embedded into matrices, i.e., document-term matrix or topic-term matrix, we can reformulate the probabilistic language models in matrixwise operations. Fig. \ref{fig.matrix} illustrate the matrix representations of document-term matrix ($F$) and topic-term matrix ($\bm{\phi}$). In this section, we will describe matrix calculations to reformulate the probabilistic language models.

\subsection{Notations}
The notations we will use are listed in Table I.\\

\begin{table}[h!]
\label{tab.not}
\centering
\caption{Notations of matrix reformulations}
\begin{tabular}{|l|p{7cm}|}
\hline
Notation & Description \\
\hline
  $m$ &  Number of documents in a collection \\
  $n$ & Number of terms in a collection \\
  $k$ & Number of latent topics in a collection \\
  $\mathbf{N}_d$ & $m$-dimensional vector representing number of terms in each of $m$  documents \\
  $\bm{\mu}$ & $m$-dimensional vector in which each element is the Dirichlet prior $\mu$  \\
  $\lambda$ & A parameter between $0$ and $1$ \\
  $\mathbf{F}$ & $m\times n$ matrix in which each element represents the frequency of a term that occurs in a document \\
  $\mathbf{F}_q$ & $1\times n$ matrix in which each element represents the frequency of a term that occurs in query $q$ \\
  $\mathbf{C}$ & $n$-dimensional vector representing the frequency of each term that occurs in a collection \\
  $\mathbf{I}_m$ & $m$-dimensional vector with values equal to one \\
  $\mathbf{I}_k$ & a $k$-dimensional vector with values equal to one \\
  $\bm{\theta}$ & $m\times k$ matrix in which each element is the probability of a topic $z$ given a document, i.e., $p(z|d)$ \\
  $\bm{\phi}$ & $k\times n$ matrix in which each element is the probability of a word given a topic, i.e., $p(w|z)$ \\
  $\mathbf{P}$ & $m\times n$ document-term probability matrix calculated by the LMD model \\
  $s_{LMD}$ & $m\time 1$-dimensional ranking score vector calculated by the LMD model \\
  $\mathbf{P}_{LBDM}$ & $m\times n$ document-term probability matrix calculated by the LBDM model \\
  $s_{LBDM}$ & $m\time 1$-dimensional ranking score vector calculated by the LBDM model \\
  $W$ & $k\times n$ topic-term probability matrix \\
  $D$ & $k\times m$ topic-document probability matrix \\
  $Q$ & $k\times 1$ topic-query probability matrix \\
  $s_{LDI}$ & $m\time 1$-dimensional ranking score vector calculated by the LDI model \\
  \hline
\end{tabular}
\end{table}

\subsection{Matrix Reformulation of LMD}
Having the notations, the document-term probability matrix can be written as
\begin{equation}
\label{equ.d.t.m}
\mathbf{P} = diag(\mathbf{N}_d+\bm{\mu})^{-1}\cdot \mathbf{F} + diag(diag(\mathbf{N}_d+\bm{\mu})^{-1}\cdot \bm{\mu})\cdot \frac{1}{n} \cdot \mathbf{I}_m \cdot \mathbf{C}^{T}
\end{equation}
We note that the elements in this probability matrix is equivalent to Equ. \ref{equ.qlm} which is trivial to verify. Given a query, the ranking score vector is:
\begin{equation}
s_{LMD}=\mathbf{P} \cdot \mathbf{F}_q^T.
\end{equation}
If a set of queries are given to calculate the ranking scores for each, $\mathbf{F}_q$ and $s_{LMD}$ becomes matrices instead of vectors where each column is corresponding to a query. In programming languages that facilitates matrix calculations, the implementation codes can be simply written in one line. For example, the code to calculate the ranking scores in Matlab is: 

\footnotesize
\begin{verbatim}p = diag(N_d+mu_vector)^(-1)*F+n^(-1).*diag(diag(N_d 
\end{verbatim}
\begin{verbatim}+mu_vector)^(-1)*mu_vector)*I_m*C'
\end{verbatim}
\normalsize
where \verb"N_d, mu_vector, F, n, I_m, C" are input data. 

\subsection{Matrix Reformulation of LBDM}
Given the matrix formulation of LMD, the document-term probability matrix of the LDA-based Language Model (LBDM) can be accordingly written as:
\begin{equation}
\label{equ.lbdm.p}
\mathbf{P}_{LBDM} = \lambda \mathbf{P} + (1-\lambda) \cdot \bm{\theta} \cdot \bm{\phi}.
\end{equation}
where $\bm{\theta}$ and $\bm{\phi}$ are output of LDA. We leverage JGibbLDA package \footnote{http://jgibblda.sourceforge.net/} in the Java package "MatLM". Given a query, the document ranking score vector is defined by:
\begin{equation}
\label{equ.lbdm.s}
s_{LBDM}=\mathbf{P}_{LBDM} \cdot \mathbf{F}_q^T.
\end{equation}
Then we can retrieve a number of top documents by sorting the ranking scores in descending order.

\subsection{Matrix Reformulation of LDI}
The term and document probabilities represented in topic space in the LDI model are written as:
\begin{equation}
W = \bm{\phi} \cdot diag(\mathbf{I}_k^T \cdot \bm{\phi})^{-1},
\end{equation}
and
\begin{equation}
D=W\cdot \mathbf{F}^T \cdot diag(\mathbf{N}_d)^{-1},
\end{equation}
respectively. Similarly, the given query can be represented by:
\begin{equation}
Q=W \cdot q^T.
\end{equation}
Based on equation.\ref{equ.ldi.s}, we have the document ranking score vector:
\begin{equation}
s_{LDI}=\left< \frac{D}{\|D\|}, \frac{Q}{\|Q\|} \right>
\end{equation}

\section{Implementation}
\label{sec.ie}

We have released the Java package "MatLM" by implementing the proposed matrix reformulations of the probabilistic language models. One can download the package at: \url{https://github.com/yanshanwang/JGibbLDA-v.1.0-MatLM}. In the package, Jama\footnote{Available at: http://math.nist.gov/javanumerics/jama/} and la4j\footnote{Available at: http://la4j.org/}  packages are utilized for matrix and sparse matrix calculations, respectively since they are the prevalent Java matrix packages in the academy and industry. The released package leverages the results of the JGibbLDA package, and thus, the input data follow the format required by JGibbLDA. Moreover, Apache Maven\footnote{Available at: https://maven.apache.org/} is used to manage the project package dependencies. 

Here we use pseudo codes to illustrate Jama package utilization for matrix formulation of LBDM. First, we build document-term matrix \verb"docTermMatrix" and query-term matrix \verb"queryTermMatrix" and run JGibbLDA package to obtain parameter matrix \verb"phiMatrix" and \verb"thetaMatrix" for the MatLM package. Second, we calculate the document-term probability matrix using equation. \ref{equ.d.t.m}:

\footnotesize
\begin{verbatim}Matrix docTermProbMatrix = diagTemp.times(termDoc-
\end{verbatim}
\begin{verbatim}
Matrix).plus(this.vec2Diagonal(this.diagonalInverse
\end{verbatim}
\begin{verbatim}
(this.vec2Diagonal(Nd.plus(Mu))).times(Mu)).times(Nc)
\end{verbatim}
\begin{verbatim}
.times(Id).times(C));
\end{verbatim}
\normalsize
where \verb"Mu", \verb"Nc", \verb"Id" and \verb"C" are pre-initialized. Finally, we calculate the query-document similarity scores using equations. \ref{equ.lbdm.p} and \ref{equ.lbdm.s}:

\footnotesize
\begin{verbatim}Matrix queryDocScore = docTermProbMatrix.times(lambda)
\end{verbatim}
\begin{verbatim}
.plus(thetaMatrix.times(phiMatrix).times(1-lambda))
\end{verbatim}
\begin{verbatim}
.times(queryTermMatrix.transpose()).transpose();
\end{verbatim}
\normalsize

As shown by the pseudo codes, it is trivial to implement the LBDM models. Similarly, using the proposed matrix formulations, the LMD and LDI models are also easy to implement. On the other hand, one can modify the codes to adjust to his new methods since the implementation directly deal with the input data and parameters.

 
\section{Conclusions and Discussions}
\label{sec.cd}

In this paper, we reformulate the probabilistic language models by means of matrix methodology and report a released Java software package 'MatLM` by implementing the matrix formulations of three language models. A drawback is that the Java package 'MatLM` is limited to the power of matrix calculation packages Jama and la4j. However, 'MatLM` can be trivially modified to exploit more efficient matrix packages, such as Jama-OSGI\footnote{Available at: https://github.com/muuki88/jama-osgi}, by levering GPU computation powers. 

The matrix representation is a potential formalism for standardization of language models. A possible future direction is to optimize the parameters by applying appropriate machine learning algorithms.

\section{Acknowledgements}
The authors gratefully acknowledge the support from National Institute of Health (NIH) grant 1R01LM011934.

\bibliographystyle{IEEEtran}
\bibliography{qlm_matrix_formulation}

\end{document}